# Quantum Black Holes and Atomic Nuclei are Hollow

Roumen Tsekov

Department of Physical Chemistry, University of Sofia, 1164 Sofia, Bulgaria

The quantum Schrödinger-Newton equation is solved for a self-gravitating Bose gas at zero temperature. It is derived that the density is non-uniform and a central hollow cavity exists. On the other hand, the radial distribution of the particle momentum is uniform. The temperature effect is accounted for via the Schrödinger-Poisson-Boltzmann equation, where low and high temperature solutions are obtained. Via the Schrödinger-Yukawa equation, the analysis is extended to a strong self-interacting gas, showing that the atomic nuclei are also hollow. Hollow self-gravitating Fermi gases are described by the Thomas-Fermi equation.

Let us consider a gas of $N$ identical point particles with mass $m$. The Poisson equation describes the gravitational potential $\phi$ generated by the radial-symmetric local mass density $\rho$

$$\nabla_r^2 \phi = 4\pi G \rho \tag{1}$$

where $\nabla_r$ is the radial nabla operator. Equation (1) requires further knowledge for $\rho$ or its fundamental relationship to $\phi$. In the classical case, the self-gravitating gas, cooled to zero temperature, collapses in a mass point with mass $M = Nm$, since the gravity is an attractive force and no other interactions are considered. In this state, the mass density is a Dirac delta-function and the corresponding solution of Eq. (1) is the classical potential $\phi = -GM / r$ of Newton. Since the Schwarzschild radius $r_s \equiv 2GM / c^2$ is always larger than the zero radius of a mass point, the latter is a black hole.

Obviously, quantum mechanics will disturb the classical delta-function distribution with zero dispersion. The quantum mass distribution will be scattered and the corresponding gravitational potential will be weaker. The stationary Schrödinger equation for a single particle of the quantum gas reads

$$-\hbar^2 \nabla_r^2 \psi / 2m + m\phi\psi = \varepsilon\psi \tag{2}$$

where $\psi$ is a real wave function and $\varepsilon$ is the particle energy. One can express from Eq. (2) the gravitational energy $m\phi = \varepsilon - Q$, where $Q \equiv -\hbar^2 \nabla_r^2 \psi / 2m\psi$ is the Bohm[1] quantum potential. Assuming uncorrelated bosons, condensed on the ground state, the local mass density $\rho = M\psi^2$ is a product of the mass of the gas and the probability density to find a particle at a given place. Substituting $m\phi$ and $\rho$ in Eq. (1) yields the self-consistent Schrödinger-Newton (or Schrödinger-Poisson) equation for the wave function of the self-gravitating Bose gas[2-4]

$$-\nabla_r^2 Q = 4\pi GMm\psi^2 \tag{3}$$

Although Eq. (3) is fundamentally nonlinear, it possesses is a very simple solution[3]

$$\psi = (r_0 / r)^2 / (4\pi r_0^3)^{1/2} \tag{4}$$

where $r_0 \equiv 2\hbar^2 / GMm^2$ is the gravitational Bohr radius. Compare the latter to expressions from the theory of atom, one can see the electrostatic parallel: $1/4\pi\varepsilon_0 \to G$, $e \to m$ and $Z \to N/2$. Thus, the gravitational fine-structure constant reads $m^2 / m_P^2$, where $m_P \equiv (\hbar c / G)^{1/2}$ is the Planck mass. Knowing the wave function $\psi$ in the coordinate space allows derivation of the wave function $\chi$ in the momentum space via a standard Fourier transformation

$$\chi = (p_0 / p) / (4\pi p_0^3)^{1/2} \tag{5}$$

where $p_0 \equiv \hbar / r_0$. The particle radial distribution density on absolute momenta acquires the form

$$f_p \equiv 4\pi p^2 \chi^2 = H(p_0 - p) / p_0 \tag{6}$$

where the Heaviside step function $H$ is introduce for normalization. This uniform distribution shows equal number of particles in any spherical momentum layer up to the maximal possible momentum $p_0 = GMm^2 / 2\hbar$, which is proportional to the number of particles in the gas.

Expressing the mass density from Eq. (4), substituting it in Eq. (1) and integrating twice along the distance yields the quantum gravitational potential of the condensed bosonic gas

$$\phi = -GM / r + (\hbar / mr)^2 \qquad (7)$$

As expected, $\phi$ tends to the classical Newton potential at large distance. At short distance, however, a quantum repulsion appears to compensate the quantum potential. It looks like an effective centrifugal potential with azimuthal quantum number $l = 1$ and describes gravitation in the Kerr[5] metric. However, this is not orbital rotation of the particles, since the used Laplace operator is radial and no angular momentum operator exists. Maybe some collective superfluid rotons appear in the Bose-Einstein condensate. The competition between the two terms in Eq. (7) leads to a potential minimum at $r_0$. Introducing the depth $\phi_0 \equiv \phi(r_0) = -GM / 2r_0$ of the potential well, being half of the Newton potential there, Eq. (7) can be rewritten in the dimensionless form $\phi = \phi_0 (r_0 / r)(2 - r_0 / r)$, which is presented in Fig. 1

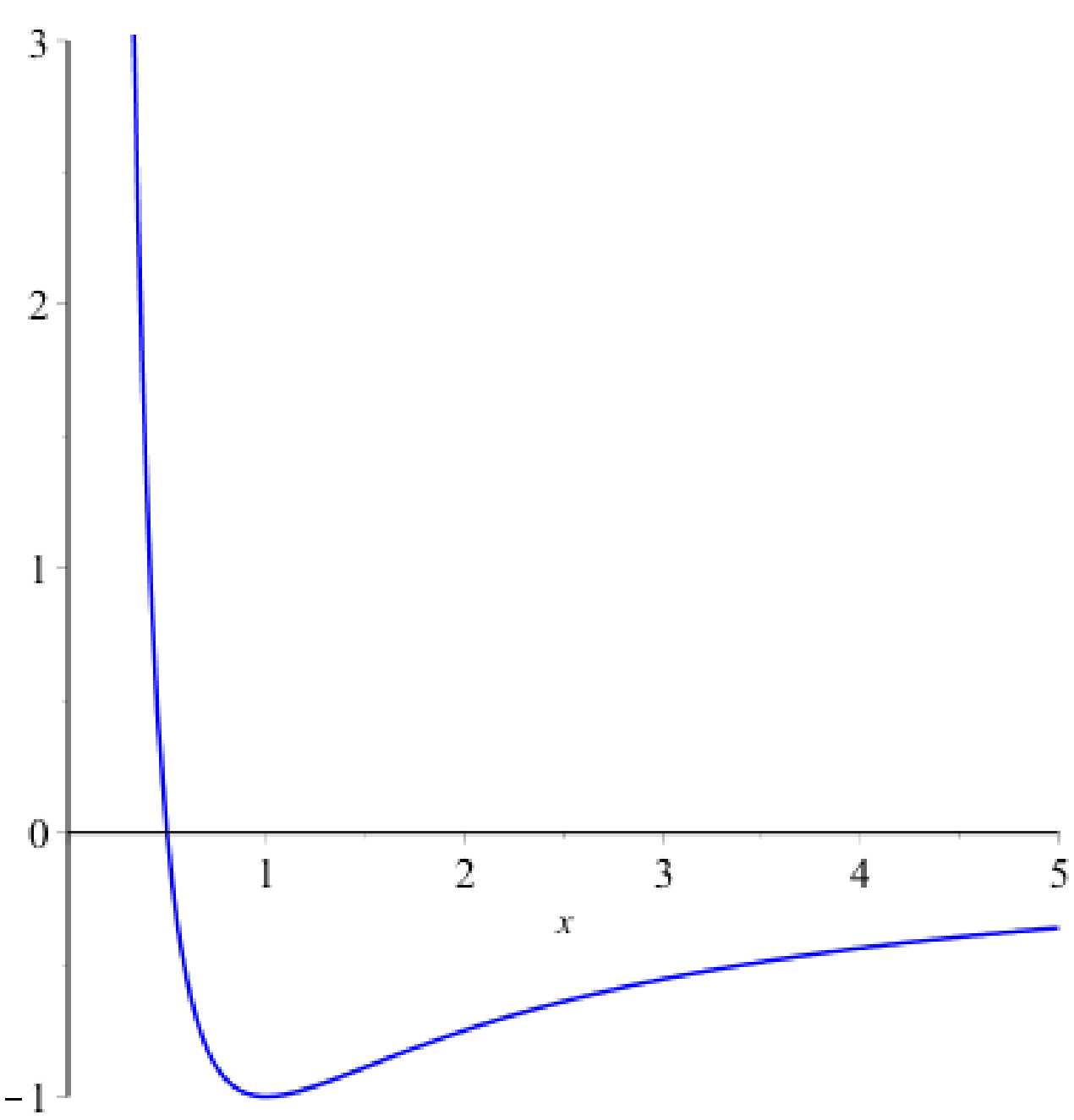

**Fig. 1.** The dimensionless potential $-\phi / \phi_0$ as a function of the dimensionless $x \equiv r / r_0$

Since the force is repulsive inside the central sphere with radius $r_0$, it is straightforward to recognize that this cavity is empty, i.e. $\psi(r < r_0) = 0$. This is also evident from the fact that the probability density of the self-gravitating gas is entirely distributed outside the cavity, since

$$\int_{r_0}^{\infty} \psi^2 4\pi r^2 dr = 1 \qquad (8)$$

Hence, the radial distribution density in the coordinate space can be further elaborated to

$$f_r \equiv 4\pi r^2\psi^2 = H(r-r_0)r_0 / r^2 \qquad\qquad M_r / M \equiv \int_0^r f_r dr = H(r-r_0)(1-r_0/r) \tag{9}$$

It is plotted in Fig. 2. The hollow cavity is due to the Heisenberg principle, since $r_0 p_0 = \hbar$, and the particles at the cavity surface possess the maximal momentum. Within the cavity, the quantum singularity overcomes the gravitational one and repels the particles outside. As expected, the probability density $f_r$ tends to a delta-function in the classical limit $\hbar \to 0$, since $r_0 \to 0$.

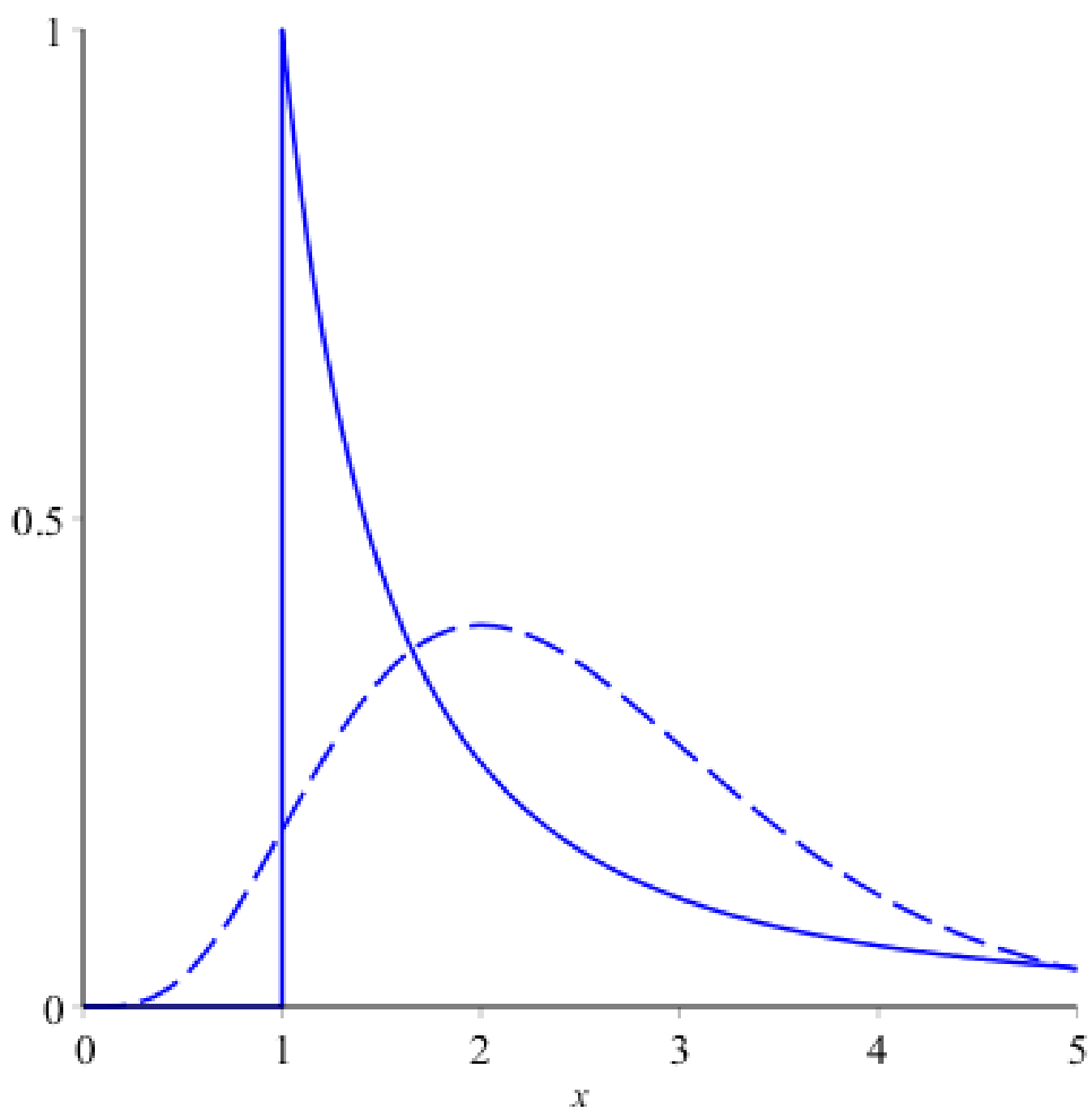

**Fig. 2.** The dimensionless radial density $r_0 f_r$ as a function of the dimensionless $x \equiv r / r_0$ from Eq. (9) (solid) and the approximation $4x^4 \exp(-2x)/3$ (dashed)

It is interesting to calculate some average characteristics of the quantum self-gravitating gas. Using the distributions (6) and (9), the mean kinetic and potential energies of a particle read

$$\varepsilon_k = \int_0^\infty (p^2 / 2m) f_p dp = p_0^2 / 6m \qquad\qquad \varepsilon_p = \int_0^\infty m\phi f_r dr = -GMm / 3r_0 \tag{10}$$

Hence, the total Bohr energy $\varepsilon = \varepsilon_k + \varepsilon_p = -p_0^2 / 2m = m\phi_0 / 2$ is negative and proportional to the square of the number of particles. At zero temperature, the bosons condensate on this ground state. It is easy to prove the virial theorem $2\varepsilon_k = \int_0^\infty (r\partial_r \phi) f_r dr$ but note that the usual expression $2\varepsilon_k \neq -\varepsilon_p$ fails, due to the quadratic quantum term in Eq. (7). Another interesting quantity is the local quantum stress tensor, defined via $\nabla \cdot \sigma = -\rho \nabla Q$,

$$\sigma \equiv (\hbar / 2m)^2 \rho \nabla\nabla \ln \rho = (\hbar / mr)^2 (2\mathbf{rr} / r^2 - \mathbf{I})\rho \tag{11}$$

The corresponding pressure $P \equiv -\mathrm{tr}(\sigma)/3 = (\hbar / 6\pi G m)(2\pi G\rho)^{3/2}$ obeys a polytropic equation of state and satisfies the standard relation $\int_0^\infty P 4\pi r^2 dr = 2N\varepsilon_k / 3$. Knowing it, one can calculate the speed of sound $\mathrm{v}_s^2 \equiv \partial_\rho P = (\hbar / 2m)(2\pi G\rho)^{1/2}$, which equals to $p_0 / 2^{1/2} m$ at the cavity surface.

The existence of cut-offs in the model is physically justified but it creates mathematical uncertainty, when a differentiation comes into use. For instance, neither of the following two well-known integrals provides the correct kinetic energy

$$(\hbar^2 / 2m)\int_{r_0}^{\infty} (\nabla_r \psi)^2 4\pi r^2 dr = 2p_0^2 / 3m \qquad -(\hbar^2 / 2m)\int_{r_0}^{\infty} (\psi \nabla_r^2 \psi) 4\pi r^2 dr = -p_0^2 / 3m$$

Moreover, they are not equal and the last integral is even negative. The correct kinetic energy equals to the average of these integrals. To clear up doubts, let us solve the Schrödinger equation (2) in the field of the unrestricted potential (7). The solution $\psi = (r / r_0)\exp(-r / r_0) / (3\pi r_0^3)^{1/2}$ is not peculiar and the kinetic, potential and total energies coincide with the estimations above. Since $\psi(r = 0) = 0$, this approximate solution confirms also that the self-gravitating Bose gas is hollow. The comparison of the corresponding radial distribution density with that from Eq. (9) is shown in Fig. (2). Of course, the approximate wave function does not satisfy Eq. (1) for the potential (7). A self-consistent procedure requires further determination of a new potential via integration of Eq. (1) and so on.

The present quantum gravitational theory is non-relativistic and it will meet the Einstein relativity theory at large mass. The Schwarzschild radius is inside the hollow cavity for small $N$. A black hole can appear at large $N$, where $r_s \geq r_0$ defines the critical mass $M^* \equiv m_P^2 / m$ of Kaup. The corresponding critical velocity $p_0^* / m = c / 2$ equals to half of the speed of light, while the critical radius of the hollow cavity $r_0^* = 2\hbar / mc$ is twice the reduced Compton wavelength of a particle. The critical number of particles $N^* = m_P^2 / m^2$ for a bosonic black hole formation is inversely proportional to the gravitational fine-structure constant. For instance, two Planck particles, either bosons or fermions, form a binary black hole and the corresponding radius of the hollow cavity is the Planck length $l_P \equiv (G\hbar / c^3)^{1/2}$. In contrast to the classical theory at $T = 0$, where any self-gravitating gas collapses to a mass point black hole, if no other interactions are present, the quantum theory above states that black holes appear only if the mass $M \geq m_P^2 / m$ is larger than a critical mass[2,3]. Figure 2 shows also that the mass distribution in the quantum black hole is not uniform. The central part of the black hole is empty and the radius of the hollow cavity is smaller than the doubled reduced Compton wavelength. As an example, let us consider an electron-positron neutral plasma. The advantage of these particles is that their masses possess purely electrostatic origin. It is expected that neutral Cooper electron-positron pairs are generated, promoted by electrostatics. Thus avoiding the fermionic repulsion, this Bose gas can easily condensate at zero temperature. Since the Compton wavelength of an electron equals to its radius divided by the electrostatic fine-structure constant, it follows that $r_0^*$ equals to 137 electron radii. The corresponding black hole critical mass is of the order of an exagram, while its size is of the order of a picometer. One can expect further that electron-positron pairs will annihilate but this will not be a problem, since even the resultant photons cannot leave the black hole. At this stage, the latter will become a photonic black hole. Especially interesting are black photons with wavelength bellow $l_P$, which are black holes by themselves, since their mass exceeds $m_P$.

The effect of temperature can be taken into account via a density-functional modification of Eq. (2) including the local Boltzmann entropy[6]

$$Q + m\phi = \mu - k_B T \ln \rho \tag{12}$$

Here $\mu$ is the constant chemical potential. This equation implies that the temperature $T$ belongs to the gas itself, not to an environment. Expressing the gravitational energy $m\phi = \mu - Q - k_B T \ln \rho$ and introducing it in Eq. (1) yields the Schrödinger-Poisson-Boltzmann equation

$$-\nabla_r^2 (Q + k_B T \ln \psi^2) = 4\pi G M m \psi^2 \tag{13}$$

At low temperature, the wave function can be written as a superposition of the zero temperature solution (4) and a small constant proportional to temperature $T$

$$\psi = (r_0 / r)^2 / (4\pi r_0^3)^{1/2} + (r_0 / \lambda_T)^2 / 2(4\pi r_0^3)^{1/2} \tag{14}$$

where $\lambda_T \equiv \hbar / (2mk_B T)^{1/2}$ is the thermal de Broglie wavelength. It is easy to check that $\psi$ satisfies Eq. (13), linearized on $T$. The gravitational potential, corresponding to Eq. (14), reads

$$\phi = -GM / r + (\hbar / mr)^2 - 2k_B T / m + (k_B T / m)[\ln(r^4 / r_0^4) - \ln(r_0^2 / \lambda_T^2)] \tag{15}$$

The last integrational constant in the brackets subtracts the thermodynamic entropy from the local one. The logarithmic potential in Eq. (15) corresponds to an additional attractive force, which shrinks the cavity radius $r_T / r_0 = 1 - r_0^2 / \lambda_T^2$. However, the potential well depth remains nearly the same $\phi_T / \phi_0 = 1 + (r_0^2 / \lambda_T^2) + (r_0^2 / \lambda_T^2)\ln(r_0 / \lambda_T)$. At large distance the potential (15) becomes positive, which suppresses the gas density there, due to the Boltzmann law. This is also evident from the radial distribution density

$$f_r \equiv 4\pi r^2 \psi^2 = H(\lambda_T - r)H(r - r_T)(r_0 / r^2 + r_0 / \lambda_T^2) \tag{16}$$

where the thermal de Broglie wave length plays the role of an upper cut-off. Therefore, the particles beyond $\lambda_T$ are not bonded to the quantum gas anymore. The mean energy can be calculated by the use of the virial theorem

$$\varepsilon = m\int_0^\infty (r\partial_r \phi / 2 + \phi) f_r dr = -p_0^2 / 2m + 3k_B T \tag{17}$$

It shows that the Dulong-Petit law holds, indicating thermal acoustic phonons, propagating with the speed of sound $v_s$. They are classical, owning to the classical Boltzmann entropy used. Due to the $T$-linearization, the formulas above are correct for $\lambda_T > r_0$, which transforms to a temperature restriction $k_B T < p_0^2 / 2m$. The temperature $T_s \equiv p_0^2 / 6mk_B$ is the sublimation point, since $\varepsilon(T_s) = 0$. At $T > T_s$ the energy becomes positive and the Bose gas evaporates completely. The quantum black hole temperature, associated usually with the temperature of the Hawking[7] radiation $T_H \equiv \hbar c^3 / 8\pi GMk_B$, must be always below $T_s$. Otherwise, the black hole is not present because $M < M^*$ follows from $T_H > T_s$ and the Schwarzschild radius is inside the hollow cavity. In critical black holes the Hawking temperature $T_H^* = mc^2 / 8\pi k_B$ touches almost the sublimation point $T_s^* = mc^2 / 24k_B$.

Motivated by the previous simple results, another simple solution $\psi = (l_T / r) / (4\pi l_T^3)^{1/2}$ of Eq. (13) is derived, clarifying the high temperature effect. Here $l_T \equiv GMm / 2k_B T$ is the gravitational Bjerrum length, which is proportional to the number of particles. As is seen, $\psi$ is completely classical, since it is valid at high temperature only. Introducing $\rho$ in Eq. (1) and integrating twice along the distance yields the gravitational potential energy

$$m\phi = k_B T \ln(r^2 / l_T^2) \tag{18}$$

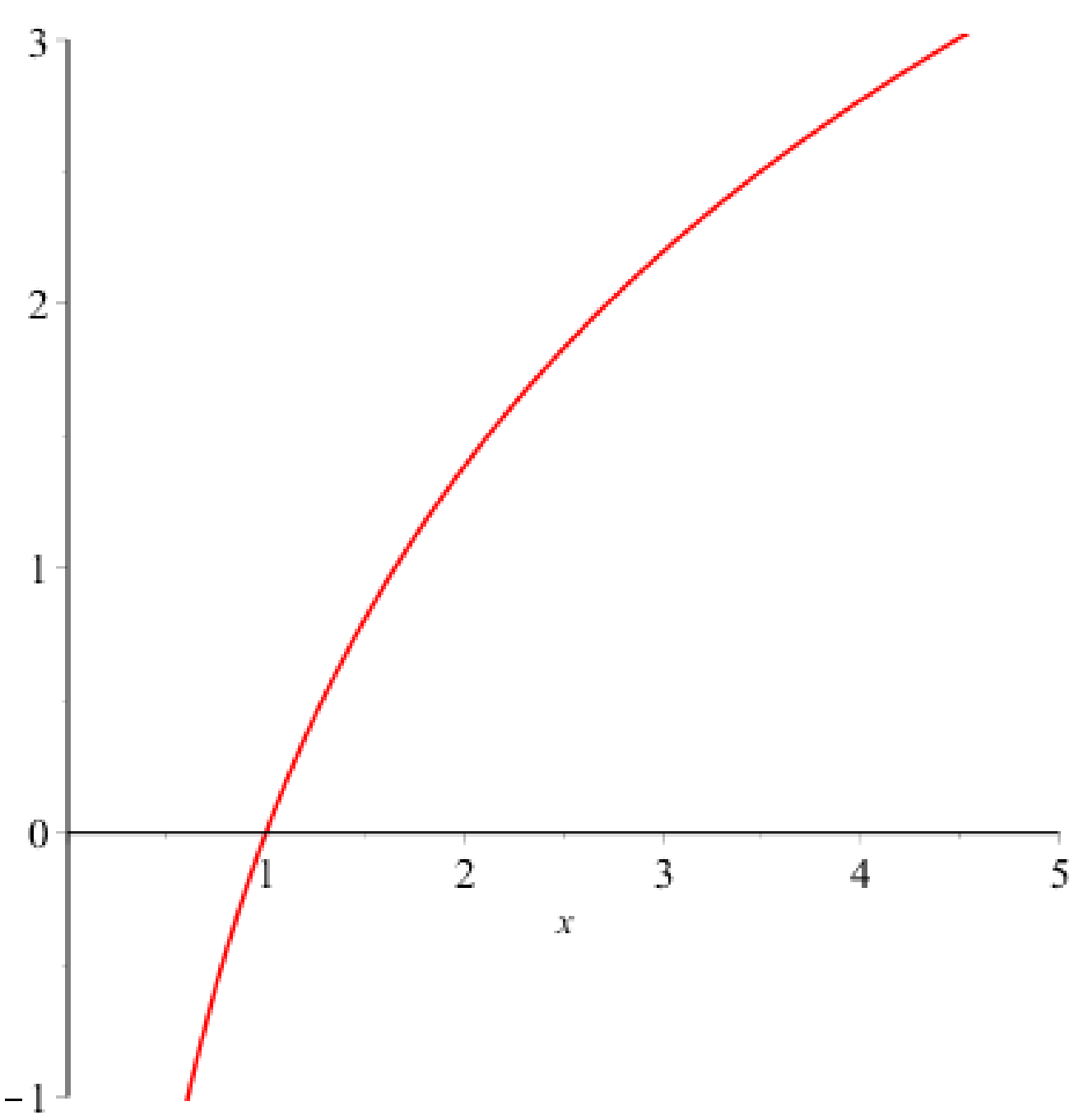

**Fig. 3.** The dimensionless potential $m\phi / k_B T$ as a function of the dimensionless $x \equiv r / l_T$

At short distance, $\phi$ is negative but weaker than the Newton potential, while at large distance it becomes positive as is shown in Fig. 3. The sign change suppresses the gas density at $r > l_T$ due to the Boltzmann law. Hence, the corresponding radial distribution density is constant up to the upper limit cut-off $l_T$

$$f_r \equiv 4\pi r^2 \psi^2 = H(l_T - r) / l_T \qquad M_r / M \equiv \int_0^r f_r dr = 1 - H(l_T - r)(1 - r / l_T) \tag{19}$$

The genetics of this uniform distribution traces back to Eq. (16), where $r_0 / \lambda_T^2 = 2 / l_T$. The corresponding kinetic, potential and total energies read $\varepsilon_k = k_B T$, $\varepsilon_p = -2k_B T$ and $\varepsilon = -k_B T$, respectively. In essence, the high temperature requires a more compacted classical gas in order the gravity to be able to keep the particles together. A possible problem of this condensed state is the stability of the proposed classical solution of Eq. (13). Thus at high temperature, it could spontaneously transform to the maximum entropy state with $\rho \to 0$ everywhere. This is not possible, however, if a black hole forms. It follows from the criterion $r_s = l_T$ that the temperature of a critical classical black hole $T^* = mc^2 / 4k_B$ is extremely high. For instance, $T^* = 1$ GK in the case of a black hole made from the already mentioned electron-positron plasma. As was discussed in the beginning, any amount of classical gas of mass points could form a black hole, but without cooling, the latter would be very hot. Moreover, cooling of a black hole is perhaps impossible, since even heat cannot leave it, unless a special kind of heat transfer is available through the event horizon.

Another interesting example is the atomic nucleus, where the protons and neutrons with mass $m$ attract each other via the Yukawa[8] potential energy $\varphi = -(\hbar^2 / 2m\lambda_Y r)\exp(-r / \lambda_Y)$. The Yukawa screening length $\lambda_Y = 1.3\,\mathrm{fm}$ coincides numerically with the Compton wavelength of a nucleon $2\pi\hbar / mc$. The corresponding screened Poisson equation reads

$$\nabla_r^2 \varphi - \varphi / \lambda_Y^2 = 2\pi\hbar^2 \rho / m^2 \lambda_Y \qquad (20)$$

Expressing $\varphi = \varepsilon - Q$ from the Schrödinger equation yields the Schrödinger-Yukawa equation[3]

$$-\nabla_r^2 Q + (Q - \varepsilon) / \lambda_Y^2 = 2\pi\hbar^2 \rho / m^2 \lambda_Y \qquad (21)$$

Far from the nucleus center $r > \lambda_Y$, the first term is negligible as compared to the second one. Thus, Eq. (21) reduces to the well-known nonlinear Gross-Pitaevskii equation[9, 10] with an attractive pseudo-potential, dependent on the particle density,

$$Q - 2\pi\hbar^2 \lambda_Y \rho / m^2 = \varepsilon \qquad (22)$$

This equation is traditionally used for bosons but it can be adapted for fermionic nucleons as well. Since the size of nucleons is commensurable with $\lambda_Y$, it is reasonable to estimate the mass density by the geometric average $\overline{\rho} \equiv 3M / 4\pi R^3$, where $R$ is the nucleus radius. Substituting it in Eq. (22) results in a Schrödinger equation, describing nucleons in a constant potential well,

$$Q - 3\hbar^2 \lambda_Y N / 2mR^3 = \varepsilon \qquad (23)$$

which is a very popular model for the atomic nuclei. The closed 3D packing implies also that the depth of the potential well is independent of the number of nucleons in the heavy nucleus. It can be estimated as 3 times the Yukawa potential energy, i.e. $-3\hbar^2 / 2m\lambda_Y^2 = -36\,\mathrm{MeV}$. Comparing this expression with Eq. (23) unveils the well-known empirical dependence $R = \lambda_Y N^{1/3}$ of the nucleus radius on the number of nucleons. Since the Fermi energy of an equal mix of protons and neutrons is $Q = \hbar^2 (9\pi N / 8R^3)^{2/3} / 2m = 28\,\mathrm{MeV}$, the binding energy amounts to 8 MeV.

On the other hand, the first term dominates Eq. (21) close to the nucleus center, where $r < \lambda_Y$. Neglecting now the second term yields the Schrödinger-Newton equation (3), where $Gm^2$ is replaced by the 37 orders of magnitude larger Yukawa magnitude $\hbar^2 / 2m\lambda_Y = \hbar c / 4\pi$

$$-\nabla_r^2 Q = \hbar c N \psi^2 \qquad (24)$$

Hence, all previous considerations apply for a nucleus near its center, as well, and confirm that the atomic nuclei are also hollow. The strong force Bohr radius $r_0 = 8\pi\hbar / Mc = 4\lambda_Y / N$ decreases also with increase of $N$. It is 4 times the Compton wavelength of the entire nucleus and represents the quantum Kerr length parameter. The lack of material constants on the right-hand side of Eq. (24) shows that the strong force is due to a super-relativistic effect, characterized simply by the product $\hbar c$ as the Casimir force. Since the nucleons are fermions, the results above are rigorous for the deuteron. Its hollow cavity radius of 2.6 fm is commensurable with the known charge radius of 2.1 fm. The binding energy of 3.0 MeV, however, is much higher than the experimental value of 1.1 MeV. Their radio $e$ shows that it is an effect of the neglected screening in Eq. (24). Therefore, for better understanding of the deuteron one should solve the Schrödinger-Yukawa equation (21). It is known that annihilation of the electron-positron Cooper pairs can create some bosonic hadrons, which interact strongly in contrast to the original leptons. The critical mass $M^* = (2mc^2 \lambda_Y / G)^{1/2}$ for an extreme Kerr black hole, formed via strong self-interacting bosons, is only 3.5 times the Planck mass. The corresponding critical radius of the hollow cavity equals to $7l_P$ and the local quantum pressure $P = 2\hbar^3 / 3m^2 c r^6$ is universal in such Bose gases.

The fact that $M^* = (4\pi\hbar c / G)^{1/2}$ is a universal constant, being independent of $m$, is due to the universality of the Yukawa interaction, as discussed before. Since the particle mass is not a parameter in the classical black hole mechanics, the latter is probably correct for matter collapsed via strong forces. Our analysis is applicable also to the internal structure of hadrons, leading to the conclusion that these composite particles are probably also hollow.

Finally, the present cut-off model is applicable to fermions as well. In a first approximation, one can imagine without proof an energy spectrum $\varepsilon_n \equiv \varepsilon / n^2$ like in the Bohr atom. Since each energy level accommodates $2n^2$ fermions, $n_F = (3N/2)^{1/3}$ and the Fermi energy equals to $E_F = \varepsilon / n_F^2$. The total energy $E = 2\varepsilon n_F$ of the entire Fermi gas defines the mean energy per particle $E / N = 3E_F$, being 3 times the negative Fermi energy. It is not clear, however, how correct these estimates are, since the bottom energy $\varepsilon$ is generated by bosons, condensed solely on their ground state. To improve the analysis, one can express from scaling considerations the quantum potential $Q = C - \hbar^2 (2\pi\nu / 3)^{2/3} / 2m$, where $\nu = \rho / m$ is the local particle density and $C$ is an unknown constant. Similar to the bosonic case, the quantum potential is attractive. Because the gravitational energy $m\phi$ is proportional to $-Q$, Eq. (1) reduces with this ansatz to the gravitational Thomas-Fermi equation[2]

$$\hbar^2 \nabla_r^2 (2\pi\nu / 3)^{2/3} / 2m = 4\pi G m^2 \nu \qquad (25)$$

Encouraged by the previous results, one is looking for a normalized particle density with a hollow central cavity in the form

$$\nu = H(r - r_0) 3N r_0^3 / 4\pi r^6 \qquad \int_0^\infty \nu \, 4\pi r^2 dr = N \qquad (26)$$

The corresponding radial probability distribution density reads $f_r \equiv 4\pi r^2 \nu / N = H(r - r_0) 3 r_0^3 / r^4$ and the mass accumulates via the cubic law $M_r / M = H(r - r_0)(1 - r_0^3 / r^3)$. It is straightforward to prove that $\nu$ from Eq. (26) is the solution of Eq. (25), if the hollow cavity radius is given by

$$r_0 \equiv \hbar^2 / G m^3 (N/2)^{1/3} \qquad (27)$$

This radius for the Fermi gas is larger than that for the Bose gas. They coincide only at $N = 2$, since there is no statistical difference between bosons and fermions in this case. Interestingly, the Fermi density decreases with distance sharper than the Bose one. However, since the hollow cavity is larger, the Bose gas cumulative mass is always higher than the Fermi one.

Knowing the density from Eq. (26), accomplished by Eq. (27), allows integration of Eq. (1) to obtain the quantum gravitational potential for the Fermi self-gravitating gas

$$\phi = -GM/r + 2(\hbar^3/2Gm^4r^2)^2 = \phi_0(r_0/r)(4 - r_0^3/r^3)/3 \qquad (28)$$

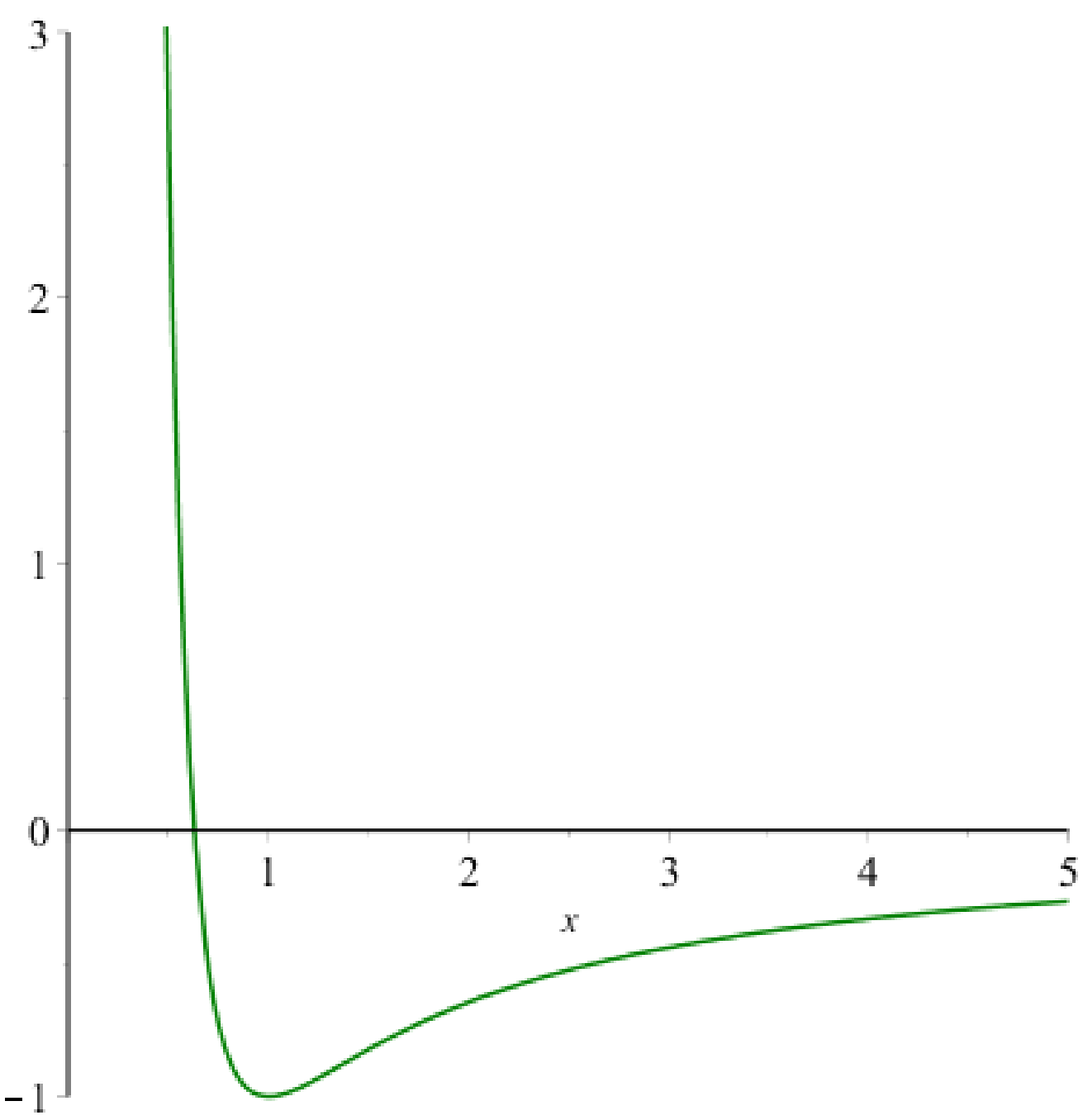

**Fig. 4.** The dimensionless potential $-\phi/\phi_0$ as a function of the dimensionless $x \equiv r/r_0$

As is seen from Fig. 4, it possesses a minimum $\phi_0 \equiv -3GM/4r_0$ at the cavity surface $r_0$, as well. Compare it with Eq. (7) confirms a stronger quantum repulsion between fermions, as expected. Using the virial theorem one can calculate the corresponding mean kinetic and potential energies of the entire Fermi gas

$$E_k = m\int_0^\infty (r\partial_r\phi/2)\nu 4\pi r^2 dr = 9GM^2/56r_0 \qquad E_p = m\int_0^\infty \phi\nu 4\pi r^2 dr = -36GM^2/56r_0$$

The unspecified constant $C = 5E_k/3N$ follows from the alternative relation $E_k = \int_0^\infty Q\nu 4\pi r^2 dr$. The total energy of the gas is negative and amounts to

$$E = E_k + E_p = -27GM^2 / 56r_0 = 27\varepsilon(N/2)^{1/3} / 7 \quad (29)$$

This binding energy is about 4/3 times the initial estimate $E = 2\varepsilon(3N/2)^{1/3}$. On the other hand, the total energy $E = N\varepsilon$ of the Bose gas decreases much sharper with the increase on the number of particles, since all the bosons condensate on the ground state. Defining the critical black hole via $r_s = r_0^*$ yields the critical number $2^{1/2} N^* = m_P^3 / m^3$ of fermions, which corresponds well to the Chandrasekhar limit. Finally, the present analysis is even better for the description of the atomic nuclei, since nucleons are fermions. Replacing $Gm^2$ by $\hbar c / 4\pi$ in Eq. (27) yields a new expression for the radius of the hollow cavity in the nuclei, $r_0 = 2\lambda_Y / (N/2)^{1/3}$. As expected, it coincides with the previous estimate for the deuteron radius. For heavier nuclei, the ratio between the cavity and nucleus radii decreases softer than bosons with increase on the number of particles. Since the nuclei consist of two kinds of fermions, nearly equal mix of protons and neutrons, the correct formula reads $r_0 / R = (2/N)^{2/3}$. For instance, this ratio is about 1/24 in Uranium, while in Oxygen the hollow cavity extends up to 1/4 of the nucleus radius. The empty volume ratio $(r_0 / R)^3 = (2/N)^2$ is 25% in Helium and 100% in Deuterium nuclei.

In general, one expects any quantum self-attracting matter of identical particles to create a hollow cavity around the mass center, because the particle indistinguishability dictates that the unique point in space at $r = 0$ remains unoccupied. The appearance of the hollow cavity solves directly the problem with the gravitational potential singularity. Moreover, it protects also the matter from a black hole collapse, as a shield over the Schwarzschild radius. Another problem of black holes is the mass defect, which will obviously reduce the strength of the crucial gravitational self-attraction. Thus, a black hole itself can collapse at the end into a zero mass object if there is a way to dissipate its energy via heat transfer through the event horizon. This could happen, for instance, by collisions of particles from the black hole and from its surrounding, which takes place on the border at $r_s$. Hence, the free particles will not enter the black hole but could exchange energy with the particles, captured already inside the black hole.

1. D. Bohm, *Phys. Rev.* **85**, 166 (1952)
2. R. Ruffini and S. Bonazzola, *Phys. Rev.* **187**, 1767 (1969)
3. M. Membrado, A. F. Pacheco and J. Sanudo, *Phys. Rev. A* **39**, 4207 (1989)
4. P. H. Chavanis, *Phys. Rev. D* **84**, 043531 (2011); *ibid.* **94**, 083007 (2016)
5. R. P. Kerr, *Phys. Rev. Lett.* **11**, 237 (1963)
6. R. Tsekov, *Int. J. Theor. Phys.* **48**, 2660 (2009)
7. S. W. Hawking, *Nature* **248**, 30 (1974).
8. H. Yukawa, *Proc. Phys. Math. Soc. Japan* **17**, 48 (1935)
9. E. P. Gross, *Il Nuovo Cimento* **20**, 454 (1961)
10. L. P. Pitaevskii, *Sov. Phys. JETP* **13**, 451 (1961)